# A METHOD OF RISK ANALYSIS AND THREAT MANAGEMENT USING ANALYTIC HIERARCHY PROCESS: AN APPLICATION TO AIR DEFENSE

Manvi Sahni and Sumanta Kumar Das

## ABSTRACT

Efficient risk analysis and threat management are essential requirements of modern air defense (*AD*) systems. The paper is a half-way between the analytic hierarchy process (*AHP*) and the practical reasoning to model and analyze the risk and threat related to military *AD* applications. The models are applied for decision making tasks of *AD* command and control (*C2*) for assessing and prioritizing the threat from hostile targets for efficient risk management. The paper presents a method for threat assessment using the fuzzy set theory, the *AHP* and the technique for order preference by similarity to ideal solution (*TOPSIS*). The target's threat attributes are first represented using the fuzzy set theory. The subjective opinions of the experts about different alternatives are quantified and ranked following the solutions of the *AHP* process. These solutions of *AHP* are obtained through the *TOPSIS* for prioritization. The models are implemented in a simulated environment. The simulated system runs without any human intervention, and represents the state-of-the-art model for the *C2* system. The use of the fuzzy set theory, *AHP* and *TOPSIS* for decision making task is particularly useful from the point of view of the futuristic risk and threat management in the battlefield. This method is easy to implement in practice and good at real-time application.

## 1. Introduction

In recent times information sharing and collaborative decision making over the defense networks have completely revolutionized the air combat operations. Today's offensive forces are equipped with sophisticated electronic attacking (*EA*) or electronic counter measuring (*ECM*) devices (for electronic jamming against radar and communications), airborne warning and controlling system (*AWACS*) aircrafts, high precision air-to-air, air-to-surface missiles, high speed fighters, bombers, unmanned air vehicles (*UAV*) etc. To respond to these, the defensive forces rely on early warning surveillance or tracking radar that has electronic counter counter measure (*ECCM*) anti jamming technologies, high-tech command and controls (*C2*) that robustly assess the threats and efficiently allocate right weapons for engaging right targets. Risk analysis and threat management of such decision making *C2* is of utmost importance to survive with such technological advancement.

## 2. Literatures Review

Usually decision making processes of *C2* involve an *OODA* (Observe-Orient-Decide-Act) loop or variants of it (Bolderheij et al., 2006). Along with the *OODA* loop recently, the *AHP* (Analytic Hierarchy Process) architectures (Saaty, 1980) of multiple attribute decision making (*MADM*) are also becoming popular because of it′s enhanced capability of practical reasoning for developing intelligent systems. It has the advantages from the user perspective in terms of both speed and ease of development of models.

Any air defense (*AD*) system is highly dependent on classifying targets, doing intent recognition, threat assessment (*TA*) and weapon allocation (*WA*). Several multidisciplinary studies have been performed to solve such problems. The *MADM* has been applied for threat assessment (*TA*) by Changwen and You (2002).

Looking at the real applications of *AHP* technologies starting from the research proposal evaluation (Beynon, 2005) to military resource allocation to examine judgment consistency (Jeonghwan et al. 2010) or batch plant design (Aguilar, 2009), one can think of applying these technologies to *C2* processes of *AD* system.

The Technique for Order Preference by Similarity to Ideal Solution **(***TOPSIS*), known as one of the most classical *MADM* methods, was first developed by Hwang and Yoon (1981), is based on the idea that the chosen alternative should have the shortest distance from the positive ideal solution (*PIS*) and on the other side the farthest distance of the negative ideal solution (*NIS*). Abo-sinna and Amer (2005) extend *TOPSIS* approach to solve multi-objective nonlinear programming problems. Jahanshahloo et al. (2005) extends the concept of *TOPSIS* to develop a methodology for solving *MADM* problems with interval data.

## 3. Objectives

The paper is mainly focused on two aspects, firstly on the modeling the *C2* of *AD* system in terms of *AHP* architectures, and secondly, evaluating the system on the basis of correct decisions in a simulated environment and by the opinion of human operators. The proposed approach is based on two-step *AHP* and *TOPSIS* methodology for prioritizing the threats. The problem is designed as single participant *MADM*. Different targets attributes are considered as criterions and the different targets with different flying status are considered as alternatives.

## 4. Methodology

An *AHP* hierarchy of the proposed system is shown in figure 1. Different target's characteristics like range, speed, altitude, lethality, intent and angle of attack are considered as the criterions for determining the threat. Different targets such as fighter, bomber, a group of fighter, a group of bomber, electronic aircraft, airborne warning and controlling system (*AWACS*) aircrafts, etc. are considered as the different alternatives of the system. The goal or objective of the decision-making process is placed at the top level of the hierarchy. The goal or objective of the *AHP* in this work is the risk analysis and threat management. The criteria and decision alternatives come in the subsequent descending levels.

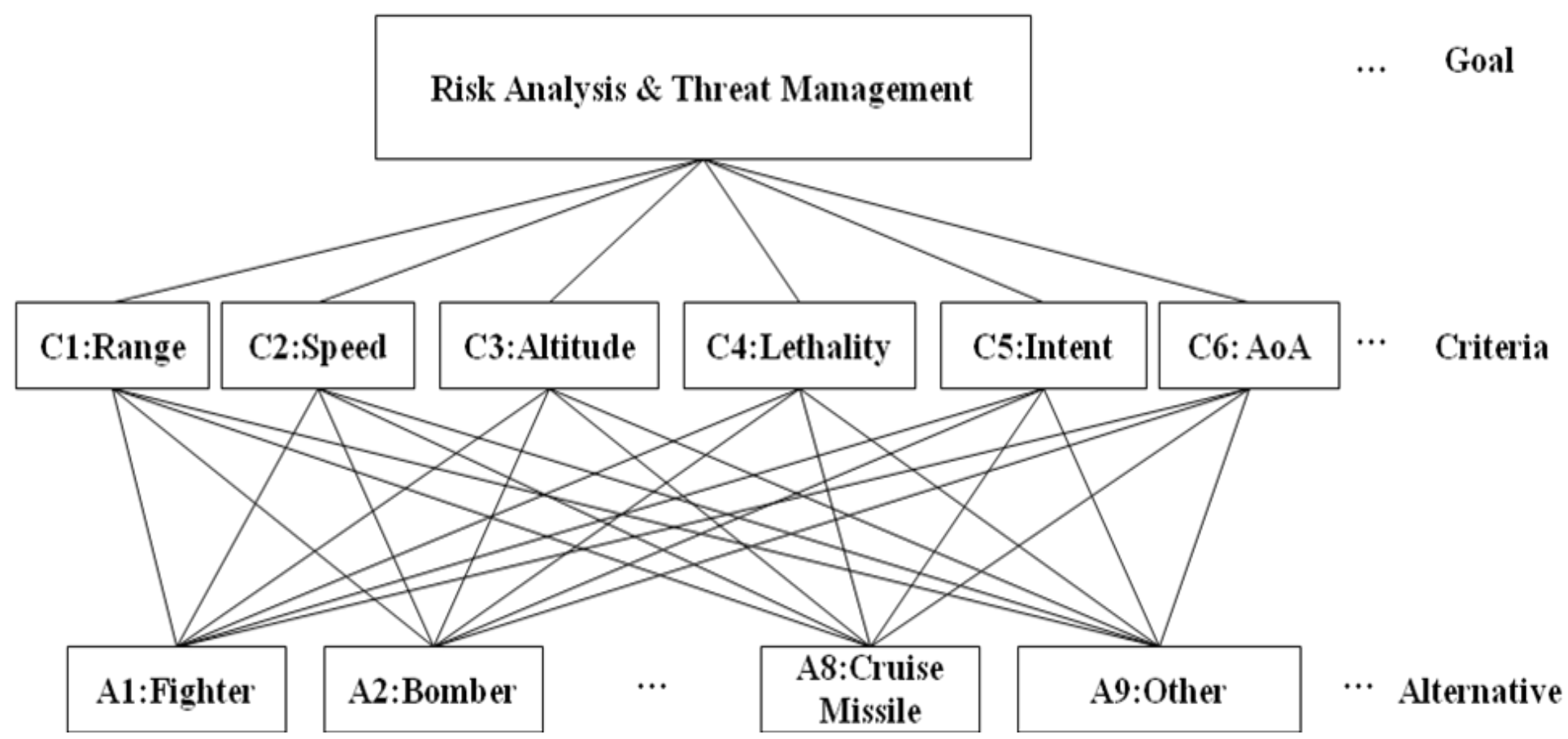

Figure 1: Hierarchy structure of AHP used

The decision maker then makes 15 pair-wise judgments among 6 criteria ((6. (5-1)/2) =15).

| | C1 | C2 | C3 | C4 | C5 | C6 | $w_i$ | Prod. | Ratio |
|---|---|---|---|---|---|---|---|---|---|
| C1:Range | 1.00 | 2.00 | 0.33 | 0.20 | 0.33 | 0.50 | 0.07 | 0.72 | 9.73 |
| C2:Speed | 0.50 | 1.00 | 0.11 | 0.11 | 0.20 | 0.33 | 0.04 | 0.42 | 9.63 |
| C3:Altitude | 3.00 | 9.00 | 1.00 | 2.00 | 2.00 | 2.00 | 0.27 | 2.69 | 9.91 |
| C4:Lethality | 5.00 | 9.00 | 0.50 | 1.00 | 2.00 | 3.00 | 0.22 | 2.17 | 10.03 |
| C5:Intent | 3.00 | 5.00 | 0.50 | 0.50 | 1.00 | 2.00 | 0.15 | 1.46 | 9.98 |
| C6:AoA | 2.00 | 3.00 | 0.50 | 0.33 | 0.50 | 1.00 | 0.11 | 1.11 | 9.94 |
| | | | | | | | | CI= | 0.07 |

For C1: Range

| C1 | A1 | A2 | A3 | A4 | A5 | A6 | A7 | A8 | A9 | $w_i$ | Prod. | Ratio |
|---|---|---|---|---|---|---|---|---|---|---|---|---|
| A1 | 1 | 2 | 1/3 | 1/5 | 1/3 | 1/2 | 2 | 3 | 2 | 0.07 | 0.72 | 9.73 |
| A2 | | 1 | 1/9 | 1/9 | 1/5 | 1/3 | 1 | 2 | 2 | 0.04 | 0.42 | 9.63 |
| A3 | | | 1 | 2 | 2 | 2 | 5 | 9 | 5 | 0.27 | 2.69 | 9.91 |
| A4 | | | | 1 | 2 | 3 | 4 | 2 | 3 | 0.22 | 2.17 | 10.03 |
| A5 | | | | | 1 | 2 | 3 | 4 | 2 | 0.15 | 1.46 | 9.98 |
| A6 | | | | | | 1 | 2 | 6 | 2 | 0.11 | 1.11 | 9.94 |
| A7 | | | | | | | 1 | 2 | 2 | 0.05 | 0.51 | 9.60 |
| A8 | | | | | | | | 1 | 2 | 0.04 | 0.39 | 9.49 |
| A9 | | | | | | | | | 1.00 | 0.04 | 0.40 | 9.55 |
| | | | | | | | | | | | CI= | 0.07 |

For C2: Speed

| C1 | A1 | A2 | A3 | A4 | A5 | A6 | A7 | A8 | A9 | $w_i$ | Prod. | Ratio |
|---|---|---|---|---|---|---|---|---|---|---|---|---|
| A1 | 1 | 3 | 2 | 3 | 5 | 5 | 2 | 3 | 3 | 0.26 | 2.73 | 10.38 |
| A2 | | 1 | 2 | 1 | 3 | 2 | 1/2 | 1 | 1 | 0.11 | 1.14 | 10.45 |
| A3 | | | 1 | 1/2 | 1/2 | 1/2 | 1 | 1/2 | 1/2 | 0.06 | 0.57 | 9.81 |
| A4 | | | | 1 | 2 | 2 | 1/2 | 1 | 1 | 0.10 | 1.04 | 10.29 |
| A5 |  |  |  |  | 1 | 1 | 2 | 2 | 2 | 0.10 | 0.99 | 10.01 |
| A6 |  |  |  |  |  | 1 | 2 | 2 | 2 | 0.10 | 1.01 | 10.01 |
| A7 |  |  |  |  |  |  | 1 | 1/2 | 1/2 | 0.09 | 0.89 | 9.82 |
| A8 |  |  |  |  |  |  |  | 1 | 1 | 0.09 | 0.87 | 9.82 |
| A9 |  |  |  |  |  |  |  |  | 1.00 | 0.09 | 0.87 | 9.82 |
|  |  |  |  |  |  |  |  |  |  |  | CI= | 0.09 |

For C3: Altitude

|  | A1 | A2 | A3 | A4 | A5 | A6 | A7 | A8 | A9 |
|---|---|---|---|---|---|---|---|---|---|
| $A_1$ (Fighter) | 1 | 7 | 1 | 5 | 1/2 | 1 | 3 | 1 | 3 |
| $A_2$ (Bomber) |  | 1 | 1/5 | 1/3 | 3 | 3 | 3 | 4 | 1 |
| $A_3$ (A group of Fighter) |  |  | 1 | 5 | 3 | 4 | 5 | 7 | 4 |
| $A_4$(A group of Bomber) |  |  |  | 1 | 5 | 5 | 7 | 9 | 3 |
| $A_5$ (Electronic Attack) |  |  |  |  | 1 | 3 | 1/3 | 1/4 | 2 |
| $A_6$ (AWACS) |  |  |  |  |  | 1 | 1/4 | 1/5 | 2 |
| $A_7$ (Tactical Ballistic Missile) |  |  |  |  |  |  | 1 | 3 | 4 |
| $A_8$ (Cruise Missile ) |  |  |  |  |  |  |  | 1 | 5 |
| $A_9$ (Other) |  |  |  |  |  |  |  |  | 1 |

For C4: Lethality

|  | A1 | A2 | A3 | A4 | A5 | A6 | A7 | A8 | A9 |
|---|---|---|---|---|---|---|---|---|---|
| $A_1$ (Fighter) | 1 | 3 | 1/5 | 1/3 | 5 | 4 | 3 | 4 | 3 |
| $A_2$ (Bomber) |  | 1 | 1/7 | 1/5 | 2 | 3 | 2 | 3 | 4 |
| $A_3$ (A group of Fighter) |  |  | 1 | 3 | 3 | 4 | 5 | 7 | 9 |
| $A_4$(A group of Bomber) |  |  |  | 1 | 5 | 5 | 3 | 3 | 4 |
| $A_5$ (Electronic Attack) |  |  |  |  | 1 | 1 | 1/3 | 1/5 | 1 |
| $A_6$ (AWACS) |  |  |  |  |  | 1 | 1/2 | 1/3 | 1 |
| $A_7$ (Tactical Ballistic Missile) |  |  |  |  |  |  | 1 | 3 | 5 |
| $A_8$ (Cruise Missile ) |  |  |  |  |  |  |  | 1 | 3 |
| $A_9$ (Other) |  |  |  |  |  |  |  |  | 1 |

For C5: Intent

|  | A1 | A2 | A3 | A4 | A5 | A6 | A7 | A8 | A9 |
|---|---|---|---|---|---|---|---|---|---|
| $A_1$ (Fighter) | 1 | 2 | 1/4 | 1/2 | 2 | 1 | 1/5 | 1/4 | 4 |
| $A_2$ (Bomber) |  | 1 | 1/9 | 1/7 | 4 | 1/3 | 1/2 | 1/3 | 2 |
| $A_3$ (A group of Fighter) |  |  | 1 | 1 | 4 | 5 | 6 | 8 | 9 |
| $A_4$(A group of Bomber) |  |  |  | 1 | 6 | 6 | 4 | 5 | 5 |
| $A_5$ (Electronic Attack) |  |  |  |  | 1 | 1 | 1/4 | 1/4 | 1 |
| $A_6$ (AWACS) |  |  |  |  |  | 1 | 1/4 | 1/4 | 1 |
| $A_7$ (Tactical Ballistic Missile) |  |  |  |  |  |  | 1 | 4 | 5 |
| $A_8$ (Cruise Missile ) |  |  |  |  |  |  |  | 1 | 4 |
| $A_9$ (Other) |  |  |  |  |  |  |  |  | 1 |

For C6: AoA

| | A1 | A2 | A3 | A4 | A5 | A6 | A7 | A8 | A9 |
|---|---|---|---|---|---|---|---|---|---|
| $A_1$ (Fighter) | 1 | 2 | 1/4 | 1/2 | 2 | 1 | 1/5 | 1/4 | 4 |
| $A_2$ (Bomber) | | 1 | 1/9 | 1/7 | 4 | 1/3 | 1/2 | 1/3 | 2 |
| $A_3$ (A group of Fighter) | | | 1 | 1 | 4 | 5 | 6 | 8 | 9 |
| $A_4$(A group of Bomber) | | | | 1 | 6 | 6 | 4 | 5 | 5 |
| $A_5$ (Electronic Attack) | | | | | 1 | 1 | 1/4 | 1/4 | 1 |
| $A_6$ (AWACS) | | | | | | 1 | 1/4 | 1/4 | 1 |
| $A_7$ (Tactical Ballistic Missile) | | | | | | | 1 | 4 | 5 |
| $A_8$ (Cruise Missile ) | | | | | | | | 1 | 4 |
| $A_9$ (Other) | | | | | | | | | 1 |

## 5. Data & Model Analysis

The training data used in this study for weight estimation are taken from fuzzy inference system (*FIS*), the details of the model can be found in Das (2014). Table 1 shows the rules used for generation of decision matrix. These rules are written on the basis of intuitive and expert considerations and then tuned by simulation tests. A Mamdani approach is followed. The input/output fuzzy sets are defined using trapezoidal and semi-trapezoidal membership functions (see figure 2 (a)). The 'and' operator and the implication methods are the product, and the defuzzification method is the weighted average. A sample decision matrix under the rules defined in Table 1 is given in the Table 2. The weights are determined using geometric mean technique (Saaty, 1996). The *TOPSIS* is used on this data set and the result is shown in Table 3. The steps of *TOPSIS* model are as follows:

- Formulation of normalized decision matrix decision matrix.(as shown in Table 1)
- Formulation of the weighted normalized decision matrix.
- Determination of the *PIS* ($C_{max}$ in Table 3) and *NIS* ($C_{min}$ in Table 3).
- Calculation of the separation measures for each alternative from the *PIS* ($S_{max}$ in Table 3) and *NIS* ($S_{min}$ in Table 3).
- Calculation of the relative closeness to the ideal solution ($G_i$ in Table 3) for each alternative.
- Prioritization of threat after ranking the $G_i$.

Table 1
Sample Fuzzy Inference Rules

| Sl. No. (Alternatives) | $C_1$: Range | $C_2$: Velocity | $C_3$: Altitude | $C_4$: AoA | $C_5$: Lethality | $C_6$: Intent | *O*: Threat |
|---|---|---|---|---|---|---|---|
| $A_1$ (Fighter) | Close | Fast | Low | High | Very Lethal | Strike | High |
| $A_2$ (Bomber) | Close | Medium | Medium | Low | Very Lethal | Bombing | High |
| $A_3$ (A group of Fighter) | Far | Fast | High | Low | Lethal | Strike | High |
| $A_4$(A group of Bomber) | Medium | Medium | Medium | High | Very Lethal | Bombing | High |
| $A_5$ (Electronic Attack) | Far | Slow | High | Low | Less Lethal | Electronic | Medium |
| $A_6$ (AWACS) | Far | Slow | High | Low | Less Lethal | Surveillance | Low |
| $A_7$ (Tactical Ballistic Missile) | Far | Fast | High | Medium | Very Lethal | Tactical bombing | Medium |
| $A_8$ (Cruise Missile ) | Medium | Medium | Low | Low | Lethal | Tactical bombing | Medium |
| $A_9$ (Other) | Slow | Slow | High | Low | Less Lethal | Surveillance | Low |

| Table 2 Sample Decision Matrix | | | | | | | |
|---|---|---|---|---|---|---|---|
| Sl. No. (Alternatives) | $C_1$: Range | $C_2$: Velocity | $C_3$: Altitude | $C_4$: AoA | $C_5$: Lethality | $C_6$: Intent | *O*: Threat |
| $A_1$ (Fighter) | 0.0896 | 0.0858 | 0.0974 | 0.1100 | 0.0922 | 0.1407 | 0.0936 |
| $A_2$ (Bomber) | 0.1226 | 0.1081 | 0.0845 | 0.0997 | 0.0859 | 0.0889 | 0.1135 |
| $A_3$ (A group of Fighter) | 0.0818 | 0.1286 | 0.0960 | 0.1085 | 0.1574 | 0.0948 | 0.0794 |
| $A_4$(A group of Bomber) | 0.1509 | 0.0875 | 0.1375 | 0.1246 | 0.1256 | 0.1304 | 0.1078 |
| $A_5$ (Electronic Attack) | 0.0833 | 0.0875 | 0.0917 | 0.1056 | 0.1463 | 0.1333 | 0.1021 |
| $A_6$ (*AWACS*) | 0.1006 | 0.0995 | 0.1347 | 0.1378 | 0.0938 | 0.1081 | 0.1234 |
| $A_7$ (Tactical Ballistic Missile) | 0.1195 | 0.1475 | 0.1304 | 0.0762 | 0.0938 | 0.0948 | 0.0993 |
| $A_8$ (Cruise Missile ) | 0.1368 | 0.1046 | 0.1160 | 0.1129 | 0.0954 | 0.0815 | 0.1418 |
| $A_9$ (Other) | 0.1148 | 0.1509 | 0.1117 | 0.1246 | 0.1097 | 0.1274 | 0.1390 |
| Weights | 0.2076 | 0.1478 | 0.1223 | 0.3842 | 0.1296 | 0.0085 | |

| Table 3 *TOPSIS* Calculation and threat prioritization | | | | | | |
|---|---|---|---|---|---|---|
| Sl. No. (Alternatives) | $C_{max}$ | $C_{min}$ | $S_{max}$ | $S_{min}$ | $G_i = S_{min} / (S_{max}+S_{min})$ | Priority |
| $A_1$ (Fighter) | 0.0422 | 0.0012 | 0.0050 | 0.0024 | 0.3217 | 8 |
| $A_2$ (Bomber) | 0.0383 | 0.0008 | 0.0036 | 0.0025 | 0.4055 | 2 |
| $A_3$ (A group of Fighter) | 0.0417 | 0.0008 | 0.0041 | 0.0028 | 0.4004 | 4 |
| $A_4$(A group of Bomber) | 0.0479 | 0.0011 | 0.0056 | 0.0037 | 0.3971 | 5 |
| $A_5$ (Electronic Attack) | 0.0406 | 0.0011 | 0.0042 | 0.0024 | 0.3620 | 7 |
| $A_6$ (*AWACS*) | 0.0529 | 0.0009 | 0.0082 | 0.0037 | 0.3091 | 9 |
| $A_7$ (Tactical Ballistic Missile) | 0.0293 | 0.0008 | 0.0014 | 0.0022 | 0.6167 | 1 |
| $A_8$ (Cruise Missile ) | 0.0434 | 0.0007 | 0.0046 | 0.0031 | 0.4027 | 3 |
| $A_9$ (Other) | 0.0479 | 0.0011 | 0.0057 | 0.0035 | 0.3787 | 6 |

The result obtained from the *TOPSIS* and the *FIS* are compared on another set of testing data and found that the proposed method performs satisfactorily.

## 6. Results and Discussion

An air combat scenario of smaller scale (200 km × 200 km) is simulated where offensive force has one ground-attack aviation regiment composed of one squadron (10 aircrafts) of high speed fighter (e.g. *A-10 Thunderbolts*) and bomber (e.g. *F-117*) each, 10 air-to-surface missiles (*Maverick*), 15 cruise missiles (e.g. *Tomahawk*), 50 smart bombs, one *UAV* and one *AWACS* aircraft. The force is using electro-optical jammer (like directed energy into the enemy's search radar) for *EA*. Each unit of this force is approaching from different directions (with different speeds, altitudes and ranges), simultaneously towards a vulnerable are or vulnerable point ((*VAVP)*, a runway and aircraft shelters), which is protected by one squadron of integrated *AD* system comprising of one surveillance radar (capable of *ECCM*), one tracking radar, interceptor aircrafts two batteries (each with 3 units) of long (e.g. *Patriot*), medium (e.g. Hawk *XXI*) and small (e.g. *NASAMS*) range *SAMs* and Anti-Aircraft Artillery and one *AHP* based *C2* system.

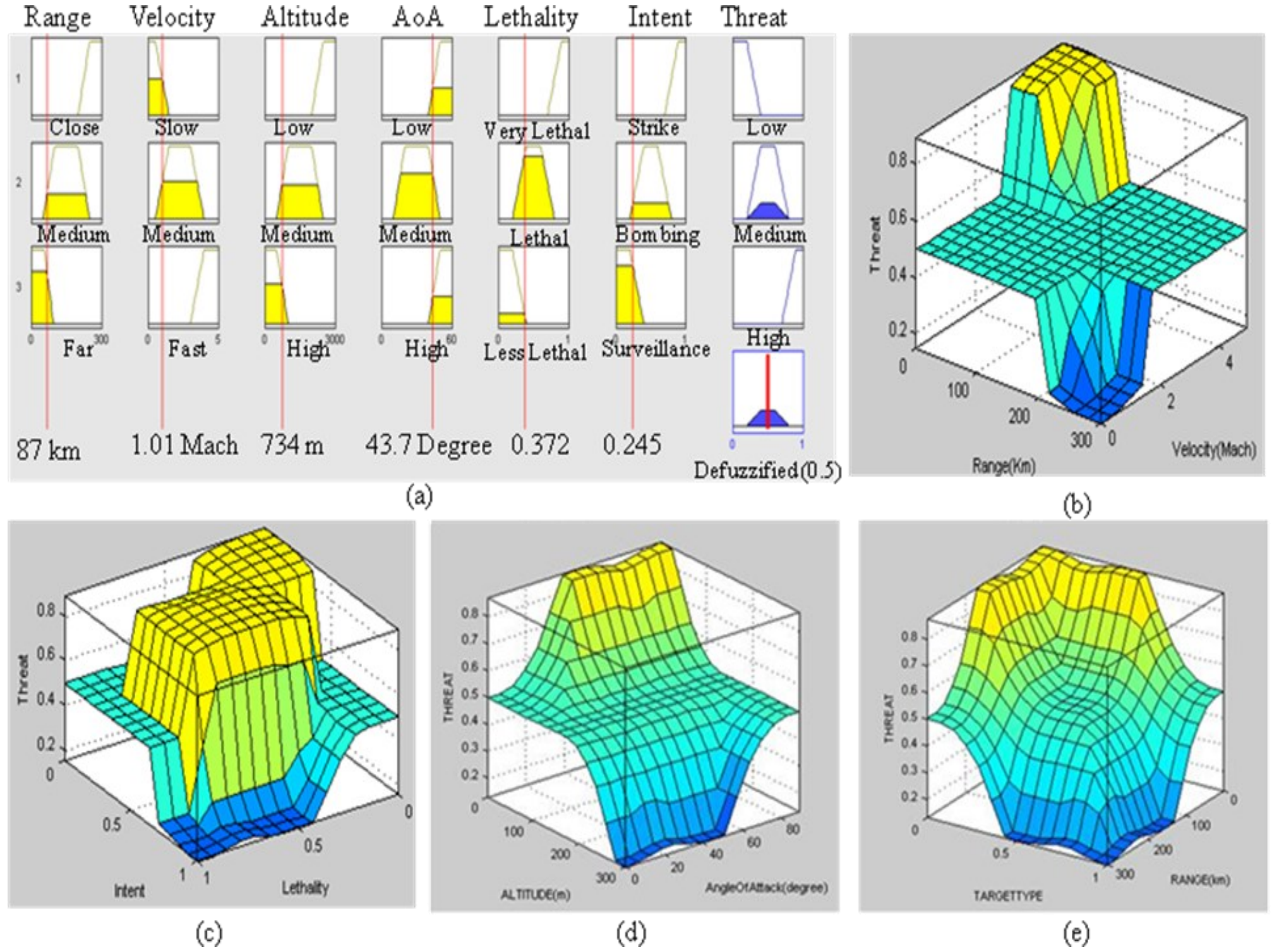

Figure 2. (a) Fuzzy inference system used for determining the decision matrix. Surface plots of threat as a function of (b) *range* and *velocity* (c) *intent* and *target types* (d) *altitude* and a*ngle of attack* and (e) *target type* and *range,* keeping other factors as fixed variable from the proposed system.

The *C2* starts prioritizing once the targets reach within 200 km range from *VAVP*. Principal findings of the simulation results suggest that if fast moving very lethal target type (a group of fighter *A-10 Thunderbolts* with speed 2.5 Mach) is very close (within 100 km) to the *VAVP*, its priority is very high as compared to a relatively slow moving target (*Tomahawk* missile with speed 0.7 Mach) which is quite far (beyond 200 km) (Fig 2(b)). Also if a lethal target (a group of bomber *F-117*) is coming with strike intention then its priority is more than a relatively less lethal target (*EA* aircraft) is coming with reconnaissance intention (Fig 2(c)). Also a target in a very low altitude (*Su-27* in a *SEAD* (Suppression of Enemy *AD*) mission) and high angle of attack is very dangerous than a target in high altitude moving in low angle of attack (*UAV*) (Fig 2(d)). Similarly, the threat of a low lethal target type (cargo aircraft) with the intention of attacking the *VAVP* (asymmetric warfare) at a very close distance is very high than a very lethal target at far range (*Su-27*) (Fig 2(e)).

## 7. Limitations

Inclusion of soft kill or non-lethal, options like decoys, chaffs, relocation of *AD* forces, deterrence measures, jamming etc. are left for future considerations. Further tests are to be done in future using two or more *C2* system to see how they may negotiate for optimal utilization of their resources.

## 8. Conclusions

In this paper, modeling of *C2* for an *AD* system is presented using the concept of *AHP* architectures. The *C2*-system takes decisions of *TA* and *WA*. The system's logic is first formulated in the form of *AHP* architectures and then implemented using the *TOPSIS*. The behavioral patterns of the *C2* system in different simulated environments are also presented.